\documentclass[pra,twocolumn,superscriptaddress,showpacs]{revtex4}
\usepackage{graphicx}
\usepackage{amsmath}
\usepackage{amssymb,bm}
\usepackage{bm,hyperref}
\usepackage{CJK}

\begin{document}

\begin{CJK*}{UTF8}{gbsn}
\title{Extended Bose-Hubbard model with pair tunneling: spontaneous symmetry breaking, effective ground state and fragmentation}
\author{Qizhong Zhu(朱起忠)}
\affiliation{International Center for Quantum Materials, School of Physics, Peking University,  Beijing 100871, China}
\author{Qi Zhang(张起)}
\affiliation{College of Science, Zhejiang University of Technology,  Hangzhou 310014, China}
\author{Biao Wu(吴飙)}
\affiliation{International Center for Quantum Materials, School of Physics, Peking University,  Beijing 100871, China}
\affiliation{Collaborative Innovation Center of Quantum Matter, Beijing 100871, China}
\date{\today}

\begin{abstract}
The extended Bose-Hubbard model for a double-well potential with pair tunneling is studied through
both exact diagonalization and mean field theory (MFT). When pair tunneling is strong enough,
the ground state wavefunction predicted by the MFT is complex and doubly degenerate
while the quantum ground state wavefunction is always real and unique. 
The time reversal symmetry is  spontaneously broken when the system transfers
from the quantum ground state into one of the mean field ground states upon a small 
perturbation. As the gap between the lowest two levels decreases exponentially with particle number, 
the required perturbation inducing the spontaneous symmetry breaking (SSB) is infinitesimal 
for particle number of typical cold atom systems. The quantum ground state is further analyzed 
with the Penrose-Onsager criterion, and is found to be a fragmented condensate. The state 
also develops the pair correlation and has non-vanishing
 \textit{pair order parameter} instead of the conventional single particle order parameter.
 When this model is generalized to optical lattice, a pair superfluid can be generated.  
The mean field ground state can be regarded as effective ground state in this simple model. 
The detailed computation for  this model enables us to offer an in-depth discussion 
of the relation between SSB and effective ground state, giving a glimpse on how nonlinearity 
arises in the SSB of a quantum system. 
\end{abstract}
\maketitle
\end{CJK*}

\section{introduction}
Bose-Hubbard model for a double-well potential has been extensively studied since the experimental realization
of Bose-Einstein condensate. Rich physics has been explored with this simple model, including the Josephson effect \cite{josephson1,josephson2} and self-trapping phenomenon for the attractive interaction \cite{trapping,liu}. Furthermore, this model is a prototype of the Bose-Hubbard model in periodic potentials, and can offer clues to   phenomenon like superfluid to Mott insulator transition that 
occurs in the thermodynamic limit \cite{SFMI}. Due to its simplicity,  one can carry out detailed 
and systematic studies of this two-site Bose-Hubbard model and obtain insights into many intriguing
phenomena. 

Here we study the Bose-Hubbard model for a double-well potential with pair tunneling, 
and show how the strong pair tunneling can change the underlying physics. 
({\it i}) When the pair tunneling effect is strong enough, the mean field ground state becomes doubly degenerate and each breaks the time reversal symmetry. In contrast, the wavefunction of the quantum ground state is always real and unique. Actually, the quantum wave wavefunction has large and equal overlap with both mean field wavefucntions. This is an analog of the NOON state  in the phase space \cite{noon}. 
({\it ii}) The model shows a general feature of a class of spontaneous symmetry breaking: the quantum ground state is unique while the effective (or mean field) ground state is degenerate. The onset of degeneracy in the effective ground state is accompanied by the appearance of quasi-degeneracy in the lowest quantum energy levels, where the energy gap decreases exponentially with particle number. The quantum ground state is unstable against small perturbations which mix up these quasi-degenerate levels. In experiments 
we always observe the effective ground state instead of the quantum one. 
Once one symmetry breaking state is chosen, the system needs an infinitely 
long time to restore the symmetry in the thermodynamic limit. 
({\it iii}) The quantum ground state of this model is a fragmented condensate, corresponding to the superposition of two coherent simple condensates. It can be characterized by a  {\it pair order parameter} instead of the single particle parameter. The extended  Bose-Hubbard model can be generalized to optical lattice; in this case a pair superfluid \cite{ps1,ps2} can be generated by a certain perturbation. 

In many condensed matter  systems the quantum ground state  can not be observed in experiments. 
What is observed is the effective ground state whose energy is almost identical to its
quantum counterpart while whose wave function has a finite difference from its quantum counterpart \cite{wezel,anderson}. The mean field ground states discussed here are  the effective ground states. 
This simple Hubbard model allows us to offer a detailed comparison between the quantum ground state 
and the effective ground state,  and discuss SSB in the perspective of effective ground state. 
We observe that there are two types of SSB: ({\it i}) The quantum ground state is degenerate. ({\it ii})
The quantum ground state is unique while the effective ground state is degenerate. The 
second type can be further divided into two subgroups due to the origin of the degeneracy. 
This may offer a fresh perspective into a question asked by 
Wen on a website,  ``What is spontaneous symmetry breaking in QUANTUM systems?"~\cite{wen}.  

The paper is organized as follows. In Sec. \ref{groundstate}, we first 
study in detail the quantum and classical ground states of the extended 
Bose-Hubbard model for a double-well potential through both exact diagonalization and MFT. 
We focus on the connection 
between these two ground states under different pair tunneling strength.
In Sec. \ref{SSB}, we discuss the relation between SSB and effective ground state. We show the feature of quantum energy levels when SSB occurs. The time scale is also determined for a symmetry breaking state to restore the symmetry. Then in Sec. \ref{fragmentation}, we use the Penrose-Onsager 
criterion to analyze the quantum ground state,
and find that it is a fragmented condensate. We further define a pair order parameter to characterize this new condensate. Finally, we summarize our results in Sec. \ref{conclusion}.

\section{Ground state of the model}
\label{groundstate}
The extended Bose-Hubbard model for a double-well potential can be described by the following
Hamiltonian \cite{liang},
\begin{align}
\hat{H}= & -t(\hat{a}^{\dagger}\hat{b}+\hat{a}\hat{b}^{\dagger})+U_{3}(\hat{n}_{a}+\hat{n}_{b}-1)(\hat{a}^{\dagger}\hat{b}+\hat{a}\hat{b}^{\dagger}) \nonumber \\
 & +\frac{U_{0}}{2}\left[\hat{n}_{a}(\hat{n}_{a}-1)+\hat{n}_{b}(\hat{n}_{b}-1)\right]\nonumber \\
 & +(U_{1}+U_{2})\hat{n}_{a}\hat{n}_{b}+\frac{U_{2}}{2}(\hat{a}^{\dagger}\hat{a}^{\dagger}\hat{b}\hat{b}+\hat{a}\hat{a}\hat{b}^{\dagger}\hat{b}^{\dagger}),
\label{qh}
\end{align}
with $\hat{a}^{\dagger} (\hat{a})$ and $\hat{b}^{\dagger} (\hat{b})$ being 
the creation (annihilation) operators in well a and b, respectively. 
$t$ is the usual single particle tunneling, $U_0$ is the usual onsite interaction, and $U_1, U_2, U_3$ 
describe the off-site interaction. Specifically, $U_1$ and $U_2$ are the inter-well particle interaction, 
and $U_3$ is the site dependent effective tunneling. All these terms show up naturally provided that one expand the field operator in 
terms of the Wannier basis and preserve all the terms. In the standard Bose-Hubbard model, the terms
involving $U_1, U_2, U_3$ are all neglected. Here the inter-well particle interaction (the last term in Eqn. \ref{qh}) serves as an effective pair tunneling, which is the focus of this work. For ultracold dilute gases with short range interaction, the contact interaction captures the essential physics, so we assume $V(\textbf{r}_1-\textbf{r}_2)=\frac{4\pi\hbar^2 a_s}{m}\delta(\textbf{r}_1-\textbf{r}_2)$, $a_s$ being the $s$-wave scattering length, and thus $U_1=U_2$. 

Below the transition temperature and for large particle number, the quantum model can be approximated by the MFT. By replacing the creation and annihilation operators with complex numbers, 
one obtains the MFT Hamiltonian, 
\begin{equation}
H= -J(a^{*}b+ab^{*})+w_{1}|a|^{2}|b|^{2}+w_{2}(a^{*2}b^{2}+a^{2}b^{*2}),
\end{equation}
where $J=t-(N-1)U_{3}$, $w_{1}=(2U_{1}-U_{0})N$, $w_{2}=NU_{1}/2$.
The ground state of this mean field Hamiltonian has been studied in Ref.~\cite{fu}. It is convenient to introduce 
a pair of canonically conjugate variables, $\theta=\theta_{b}-\theta_{a}$ and
$s=|b|^{2}-|a|^{2}$, where $a=|a|e^{i\theta_{a}}$, $b=|b|e^{i\theta_{b}}$.  
We focus on the repulsive interaction case and tune the ratio between pair tunneling 
and single particle tunneling, $\lambda=w_{2}/J$.  
In the weak pair tunneling regime, $0<\lambda<\frac{1}{2}$, 
the ground state is the fixed point $(s,\theta)=(0,0)$, which has equal population 
and zero relative phase between the wells.
In the strong pair tunneling regime, $\lambda >\frac{1}{2}$ , 
the ground state becomes two-fold degenerate with $(s,\theta)=\left(0,\pm\arccos\left(\frac{1}{2\lambda}\right)\right)$, which has zero population imbalance but nonzero relative phase 
between the two wells. The transition between these two scenarios occurs at the critical value $\lambda_c=1/2$. 

\begin{figure}[!htb]
\includegraphics[width=9.2cm]{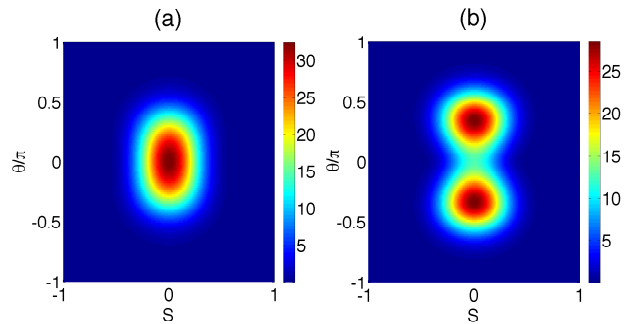}\newline
\caption{(Color online) The inner product of the quantum ground state and the mean field ground state
$|f(s,\theta)|$. (a) Weak pair tunneling regime: $U_{1}/U_{0}=0.002$ and $\lambda=0.2< \frac {1}{2}$. The mean field ground state is non-degenerate and $|f(s,\theta)|$ has only one peak.
(b) Strong pair tunneling regime: $U_{1}/U_{0}=0.02$ and $\lambda=2>\frac{1}{2}$. 
The mean field ground state is two-fold degenerate and $|f(s,\theta)|$ has two peaks.}
\label{theta}
\end{figure}

Our interest is in the connection between the quantum and mean field ground states.
The quantum ground state can be obtained by diagonalizing the quantum Hamiltonian Eqn. (\ref{qh}).
In contrast to the mean field results, the quantum ground state is always real 
and non-degenerate no matter how strong the pair tunneling is. 
To elucidate and understand this subtle discrepancy,  it is helpful to observe that 
the mean field ground state is actually the coherent state 
$(a\hat{a}^{\dagger}+b\hat{b}^{\dagger})^{N}|0\rangle$ in the second quantized language.
This allows us to  calculate the inner product  $f(s,\theta)$ between  the quantum ground state 
and the classical one. In usual cases without pair tunneling effect or weak pair tunneling regime, 
the inner product  $f(s,\theta)$ is  a function with its single peak located at  $(s,\theta)=(0,0)$ 
and its width representing the quantum fluctuations caused by interaction.
In the strong pair tunneling regime, the classical ground state becomes 
two degenerate fixed points $(0,\pm\Theta(\lambda))$,
and consequently $f(s,\theta)$ has two peaks located at those two
fixed points. This  is illustrated in Fig. \ref{theta}, where  $|f(s,\theta)|$ is plotted for a specific 
set of parameters,  $N=40$, $J=0.1$, $U_{0}=0.5$,  $U_{1}/U_{0}=0.002$, and $0.02$.  

Actually this state is the phase space analog of the self-trapping effect in a double-well potential for attractive particle interaction \cite{trapping,liu}. The self-trapping effect occurs when the attraction between particles is strong enough, and then MFT predicts that all the particles will spontaneously occupy only one well, breaking the symmetry of the double-well potential explicitly. The quantum ground state is known as the NOON state, an equal superposition of particles occupying both wells, which in fact restores the symmetry. Here in our model MFT also gives two degenerate states breaking the symmetry of the potential, and the quantum ground state can be called the NOON state in the phase space.

\section{Effective ground state and spontaneous symmetry breaking}
\label{SSB}
\begin{figure}
\begin{center}
\includegraphics[width=9cm]{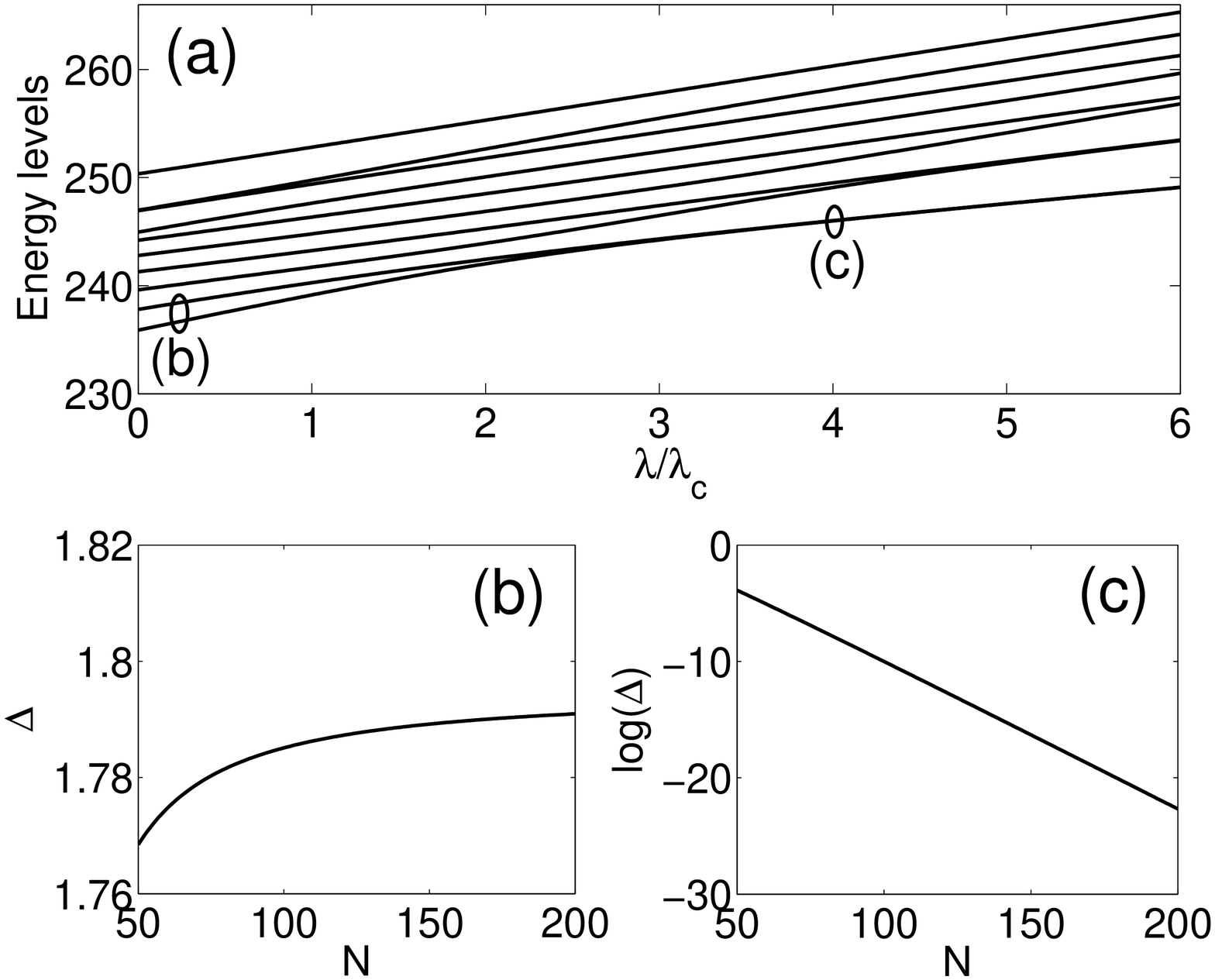}
\caption{(a) The lowest 10 energy levels as a function of pair tunneling strength, which is rescaled with respect to the critical strength $\lambda_c$. The energy spectrum shows quasi-degeneracy for strong pair tunneling. Particle number $N=50$. (b) The energy gap $\Delta$ between the lowest two states as a function of particle number $N$ at weak pair tunneling strength $\lambda/\lambda_c=0.2$ (marked by an open circle near the left-lower corner in (a)). $\Delta$ increases with $N$ before saturation. (c) $\log(\Delta)$ as a function of particle number $N$ at strong pair tunneling strength $\lambda/\lambda_c=4$ (marked by the other open circle in (a)). $\Delta$ decreases exponentially with $N$ when the pair tunneling is strong.}
\label{low10}
\end{center}
\end{figure}

Our numerical solution of the quantum model shows when the mean field ground state 
becomes degenerate, the quantum energy levels exhibit the feature of  quasi-degeneracy, where
the energy gap between the lowest two states decreases exponentially with particle number
 (see Fig. \ref{low10}).  The quasi-degeneracy indicates that the quantum ground state is unstable; very small perturbation will mix up the quasi-degenerate states. For typical cold atomic system with particle number up to $10^5\sim 10^6$, an infinitesimal perturbation will drive the system  into 
 one of the degenerate \textit{effective ground states}.

The effective ground state is associated with the experimental observability of a ground state. 
As there is always all kinds of noise in experiments, what is observed experimentally may not  
be the true ground state of the system, but the effective one. A well-known example is the gas to solid phase transition in free space \cite{wezel,anderson}. Since the center of mass momentum of the particles commutes with the whole Hamiltonian, the absolute ground state should be the eigenstate of center of mass momentum, namely, particles should be distributed homogeneously in the whole space. However, in real world we always observe a localized solid. This is because the exact ground state is unstable against infinitesimal perturbation in the thermodynamic limit, and it will spontaneously decay into a symmetry-breaking effective ground state. The quantum time crystal \cite{timecrystal} proposed by Wilczek is also such an effective ground state, whose existence needs the breaking of both translational and time
reversal symmetries. 

\begin{figure}[!htb]
\includegraphics[width=9cm]{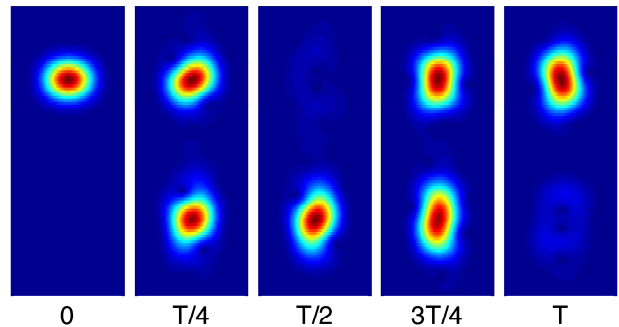}\newline
\caption{(Color online) Evolution of one of the two effective ground states under the quantum Hamiltonian in the strong pair tunneling regime. $N=50$ and $\lambda/\lambda_c=20$. The profile $|f(s,\theta)|$ has the same meaning as in Fig. \ref{theta}.}
\label{evolve}
\end{figure}

Suppose that we prepare one effective ground state. Since it is not the true ground 
state of the system, it will evolve under the quantum Hamiltonian. This process also 
determines the lifetime of the effective ground state observed in experiments.
For this model , we find that the effective ground state  evolves almost periodically 
with the period $T$ as illustrated in Fig. \ref{evolve}. As 
the energy gap $\Delta$ between the first excited state 
and the ground state decreases  with particle number $N$ as $\Delta\sim\exp(-N)$, we have
$T\sim2\pi/\Delta\sim\exp(N)$. So for large particle number, the evolution 
period is remarkably long, the system will always stay in the effective ground state 
during the course of experiments. 

From this simple model, we see clearly that the two fundamental concepts, SSB 
and effective ground state,  are closely related. We take this opportunity to 
discuss their relation.  In the perspective of effective ground state, there are  
two different types of SSB. In the first type,  the quantum ground state is degenerate; therefore, 
the effective ground state is also degenerate. At low temperature, the system falls into 
one of the quantum degenerate states, breaking the symmetry. The well-known ferromagnetic 
ground state falls into this category.  

In the second type of SSB, the quantum ground state is unique while 
the effective ground state is degenerate.  
In these systems, the degeneracy of the effective ground states arises from  
the quasi-degeneracy of the quantum energy levels: tiny perturbation can 
mix up these quasi-degenerate quantum states and generate these degenerate
effective ground states, which are observed in  experiments.
Interestingly, this category can be further divided into two subgroups as
the quasi-degeneracy has two different origins: 
({\it i}) The quasi-degeneracy of the quantum energy levels is 
caused by the large volume size of the system.  The gas-solid transition mentioned above \cite{wezel,anderson}
and the Bose-Einstein condensation of ideal gas \cite{Yukalov} are typical examples of this case. 
The degree of degeneracy in this subgroup is usually infinite. 
({\it ii}) The  quasi-degeneracy of the quantum energy levels is 
due to the interaction in the system.  For systems in this subgroup, the energy gap 
usually decreases exponentially with particle number.   
Our model belongs to this subgroup along with many other
systems reported elsewhere \cite{ueda1,ueda2,ueda3,zhai2,zhai}. 

Spontaneous symmetry breaking is a familiar and well-studied concept  in condensed matter physics.
However, it is still not fully understood. Wen recently asked, ``What is spontaneous symmetry breaking in QUANTUM systems?" on a website~\cite{wen}. This question highlights a dilemma that we all face:
on the one hand, SSB can only happen for nonlinear systems; on the other hand, quantum systems are always linear. The connection between SSB and effective ground state that we have shown 
here may provide a fresh perspective into this issue: nonlinearity needed for SSB 
is rooted in the degeneracy (or quasi-degeneracy) of quantum energy levels. 

\section{Fragmentation and pair order parameter}\label{fragmentation}
\begin{figure}[!htb]
\includegraphics[width=9cm]{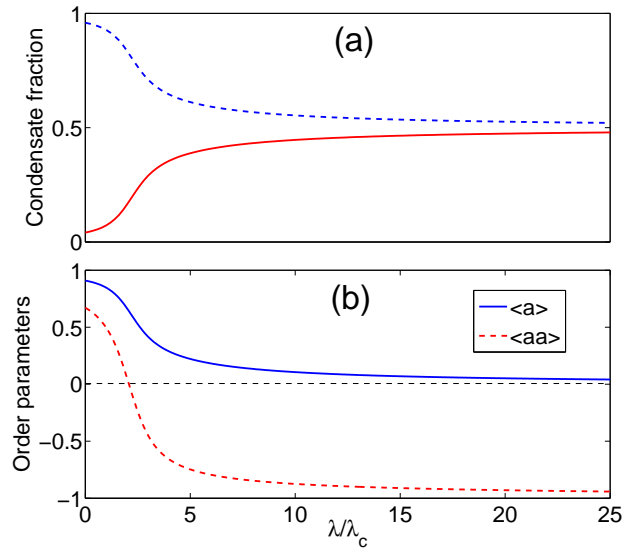}\newline
\caption{(Color online) (a) Two fragmented condensate fractions of the quantum ground state as a function of pair tunneling strength. $N=50$. (b) Order parameter $\langle\hat{a}\rangle$ (blue solid line) and pair order parameter $\langle\hat{a}\hat{a}\rangle$ (red dashed line) as a function of pair tunneling strength for the quantum ground state. They are rescaled with respect to $\sqrt{N/2}$ and $N/2$, respectively. $N=50$.}
\label{fraction}
\end{figure}

As our system is an interacting boson system,  
we use the Penrose-Onsager criterion \cite{PO} to further analyze its quantum ground state. 
In lattice models, the condensed fraction is given by
the eigenvalues of reduced single particle density matrix $\rho_1(i,j)=\langle\hat{a}_i^{\dagger}
\hat{a}_j\rangle$. Here for the double-well potential,  we have
\begin{equation}
\rho_1=\left[\begin{array}{cc}
\langle\hat{a}^{\dagger}\hat{a}\rangle & \langle\hat{a}^{\dagger}\hat{b}\rangle\\
\langle\hat{b}^{\dagger}\hat{a}\rangle & \langle\hat{b}^{\dagger}\hat{b}\rangle
\end{array}\right]\,.
\end{equation}
As shown in Fig. \ref{fraction}(a), for strong pair tunneling, both of the  eigenvalues
of  the reduced density matrix are nonzero. So,  the system is in fact a fragmented condensate.

The fragmentation of the quantum ground state corresponds to the degeneracy of mean field ground state. 
This is a general relation that is also present in other fragmented systems.
For example, in the antiferromagnetic ground state of spin-1 Bose gases \cite{pu,ho,mueller}, the quantum ground state is unique while its effective ground state has infinite degeneracy due to the intrinsic SU(2) 
symmetry. Thus the quantum ground state also corresponds to a fragmented condensate. 
Fragmented condensate is usually unstable against perturbations and will evolve into one 
simple condensate, which is one of the effective ground state. 

Distinct to the Bose-Hubbard model previously studied, one can define two kinds of order parameters for the quantum ground state in this model. One is the ordinary single particle
order parameter $\langle\hat{a}\rangle$, and the other is the pair order parameter $\langle\hat{a}\hat{a}\rangle$.
Both order parameters change with pair tunneling strength (Fig. \ref{fraction}(b)). One immediately sees that in the strong pair tunneling
limit, the single particle order parameter $\langle\hat{a}\rangle$ almost vanishes, while the pair order parameter
$\langle\hat{a}\hat{a}\rangle$ approaches a saturation value $-N/2$. This behavior indicates that for strong pair tunneling
the condensate is no longer the usual single particle condensate, but a \textit{pair condensate}. The pair order parameter changes sign during the increase of pair tunneling strength as a result of minimization of the pair tunneling energy. By tuning the ratio
of pair tunneling to single particle tunneling, the system experiences a quantum phase transition from the single particle
condensate to the pair condensate.

The pair condensate defined in this way can be extended into the case of optical lattice with periodic boundary condition, where a \textit{pair superfluid} can be defined similarly. In this lattice system, the quantum ground state is still unique while the mean field one becomes hugely degenerate due to the 
large quasi-degeneracy in the quantum energy levels.
From the analysis of the double-well model, it is clear that this pair superfluid is unstable and may decay into one of the mean field ground states, which is a simple superfluid. Among all the mean field ground states, two of them are of special interest.
Denote the phase difference between two neighboring wells as $\pm\theta$, and then there are states with consecutive $+\theta$ or $-\theta$ phase accumulation between neighboring wells, satisfying $L\theta=2q\pi$ ($L$ is the number of wells, and $q$ is the winding number of order parameter). They correspond to superfluids flowing clockwisely or counter-clockwisely. Thus these two ground states carry nonzero mass current, which is unconventional.

Note that the meaning of the pair superfluid here is different from that defined in BCS pairing or boson dimer for attractive interaction \cite{zoller,lee,ng,bonnes,chen,wessel}. Obviously no bound state or dimer is formed here. Similar order parameter also exists elsewhere \cite{zhai}, where a \textit{trion superfluid} is defined.

\begin{figure}[!tb]
\includegraphics[width=9cm]{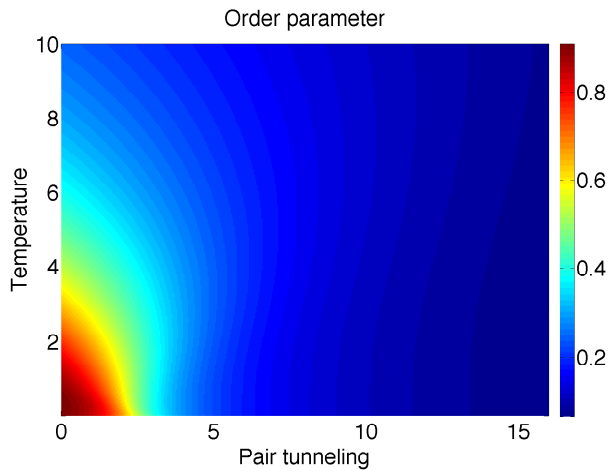}\newline
\caption{(Color online) Order parameter $|\langle\hat{a}\rangle|$ changes with temperature (in arbitrary unit) and pair tunneling strength $\lambda/\lambda_c$. $N=50$.}
\label{1op}
\end{figure}

\begin{figure}[!t]
\includegraphics[width=9cm]{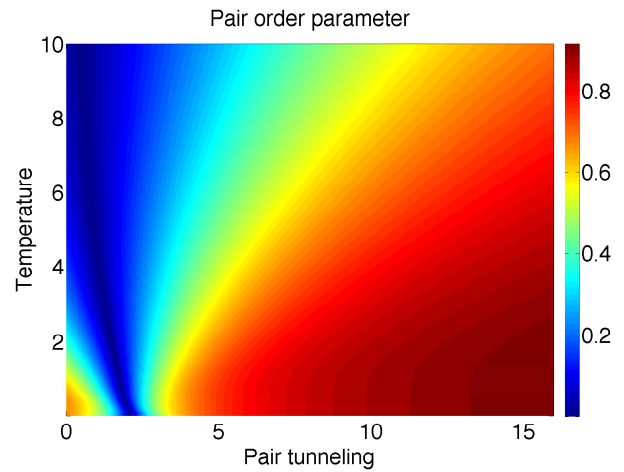}\newline
\caption{(Color online) Pair order parameter $|\langle\hat{a}\hat{a}\rangle|$ changes with temperature (in arbitrary unit) and pair tunneling strength $\lambda/\lambda_c$. $N=50$.}
\label{2op}
\end{figure}

With all the information of eigenvalues and eigenstates, one can also obtain the exact phase diagram of the double-well model at finite temperature,
as shown in Figs. \ref{1op} and \ref{2op}. At zero temperature, one can see the signature of quantum phase transition, though here due to finite
size effect, the transition becomes a crossover and the critical point also shifts. The pair condensate can also exist
at finite temperature, but is killed along with the single particle condensate at high temperature.

\section{conclusion}\label{conclusion}
In summary, we have investigated both the quantum and mean field ground states of the 
extended Bose-Hubbard model for a double-well potential with pair tunneling. 
Firstly, we find that when the pair tunneling is strong enough, 
the mean field ground state of the system  becomes complex
and two-fold degenerate, while the quantum counterpart is always real and unique.  
The quantum ground state can be regarded as a superposition of two mean field states.
Secondly, we have also discussed spontaneous symmetry breaking in the perspective of effective ground state. 
The model serves as an example where the true ground state (quantum ground state) is quite different from the effective ground state (mean field ground state) that we really observe in experiments. 
We show that the degeneracy of the effective ground state is closely related to spontaneous symmetry breaking, the quasi-degeneracy of the quantum spectrum and the stability of the quantum ground state.
For large particle number, the quantum ground state is very sensitive to
infinitesimal perturbations breaking the time reversal symmetry, resulting in one symmetry breaking effective ground state.
Finally,
we find that in the strong pair tunneling limit, the quantum ground state in fact corresponds to a fragmented condensate. In terms of the defined pair order parameter, it is also a pair condensate.
 We then provide a complete phase diagram of this model at finite temperature, and predict the instability of the pair superfluid in an optical lattice. 
 
 Although for usual systems, the pair tunneling is much weaker than the single particle tunneling,
their ratio can be tuned experimentally with the method of shaking the optical lattice \cite{graham,holthaus,grigoni,shaking}.
So the conclusion drawn here can be tested within the current experimental techniques.

\section{acknowledgement}
We would like to thank H. Zhai for
helpful discussions. Zhu and Wu are supported by the
NBRP of China (2013CB921903,2012CB921300) and the NSF of China
(11274024,11334001).  Zhang is supported by the NSF of China (11105123).

\end{document}